\documentclass[12pt]{article}
\usepackage{amssymb}
\usepackage{a4}
\begin{document}
%\sloppy \raggedbottom \setcounter{page}{1}
%%%%%%%%%%%%%%%%%%%%%%%%%%

%\newpage
%\setcounter{figure}{0} \setcounter{equation}{0}
%\setcounter{footnote}{0} \setcounter{table}{0}
%\setcounter{section}{0}

%Macro
\def\x{\mbox{x}}
% Macro for the kbar symbol:
\def\kbar{{\mathchar'26\mkern-9muk}}
%Macro for the black-board bold
\def\b#1{{\mathbb #1}}
%Macro for the calligraphic letters
\def\c#1{{\cal #1}}
%Macro
\def\nn{\nonumber \\}
%Macro
\def\1{{\bf 1}}
%Macro
\def\RH{\mbox{$\hat {\sf R}$\,}}
\def\PH{\mbox{$\sf P$}}
\def\P{\mbox{$\cal P$}}

%Macro for the universal R-matrix:
\def\R{{\cal R}\,}
%Macros for the left and right action of $U_qg$
\newcommand{\tr}{\triangleright\,}
\newcommand{\tl}{\,\triangleleft}
%Macros for the left and right action of $U_q^{op}g$
\newcommand{\tro}{\triangleright^{op}\,}
\newcommand{\tlo}{\,\stackrel{op}{\triangleleft}}
%Macros for the left and right semidirect products
\def\cross{{\triangleright\!\!\!<}}
\def\cocross{{>\!\!\!\triangleleft\,}}
%Macros for the Lie algebra and quantum group
\def\g{\mbox{\bf g\,}}
\def\A{\mbox{$\cal A$}}
\def\id{\mbox{id\,}}
\newcommand{\uot}{\underline{\otimes}}

% New commands
\newcommand{\be}{\begin{equation}}
\newcommand{\ee}{\end{equation}}
\newcommand{\bea}{\begin{eqnarray}}
\newcommand{\eea}{\end{eqnarray}}
\newcommand{\ba}{\begin{array}}
\newcommand{\ea}{\end{array}}
%


% use the thanks command within \title, \author or \address for footnotes:
% \title{Title} or  \title{Title\thanks{...}}

\title{$q$-Quaternions and deformed $su(2)$ instantons on the
quantum Euclidean space $\b{R}_q^4$\footnote{Talk given at 
the 4-th International Symposium ``Quantum Theories and Phsysics'',
 Varna, Bulgaria, August 2005. To appear in the proceedings.
Preprint 05-55 
Dip. Matematica e Applicazioni, Universit\`a di Napoli;
DSF/44-2005.}}

\author{        Gaetano Fiore, \\\\
         \and
        Dip. di Matematica e Applicazioni, Universit\`a ``Federico II''\\
        V. Claudio 21, 80125 Napoli, Italy\\
        \and
        \and
        I.N.F.N., Sezione di Napoli,\\
        Complesso MSA, V. Cintia, 80126 Napoli, Italy
        }
\date{}

\maketitle
\abstract{
We briefly report on our recent results regarding the introduction
of a notion of a $q$-quaternion and the construction of instanton
solutions of a would-be deformed $su(2)$ Yang-Mills theory on the
corresponding $SO_q(4)$-covariant quantum space. As the solutions
depend on some noncommuting parameters, this indicates that the
moduli space of a complete theory will be a noncommutative manifold.}

%%%%%%%%%%%%%%%%%%%%%%%%%%%%%%%%%%%%%%%%%%%%%%%%%%%%%%%%%%%%%
% The main text of your paper                               %
%%%%%%%%%%%%%%%%%%%%%%%%%%%%%%%%%%%%%%%%%%%%%%%%%%%%%%%%%%%%%
 \section{Introduction}

The search for instantonic solutions  has become a key point of
investigation of Yang-Mills gauge theories on noncommutative
manifolds after the discovery \cite{NekSch98}  that deforming
$\b{R}^4$ into the Moyal-Weyl noncommutative Euclidean space
$\b{R}_{\theta}^4$ regularizes the zero-size singularities of the
instanton moduli space (see e.g
\cite{SeiWit99,ConLan01,BonCicTar00,DabLanMas01,LanvSu06}). Among the
available deformations of $\b{R}^4$ there is also the
Faddeev-Reshetikhin-Takhtadjan noncommutative Euclidean space $\b{R}_q^4$
covariant under $SO_q(4)$ \cite{FadResTak89}, and it is therefore tempting to
investigate this issue on it. There is still no satisfactory formulation
\cite{JurMoeSchSchWes01} of gauge field theory on quantum group covariant
noncommutative spaces (shortly: quantum spaces) like $\b{R}_q^4$. One main
reason is the lack of a proper (i.e. cyclic) trace to define gauge invariant
observables (action, etc). Another one is the $\star$-structure of the
differential calculus, which for real $q$ is problematic. Probably a
satisfactory formulation will be possible within a generalization of
the standard framework of noncommutative geometry \cite{Con94}. Here
we leave these two issues aside and just ask for nontrivial
solutions of the deformed (anti)selfduality equations.

As known,  great simplifications in the search and classification of
instantons in Yang-Mills theory on $\b{R}^4$ occur when the latter
is promoted to the quaternion algebra $\b{H}$.  We have recently
introduced \cite{Fio05i} the notion of a $q$-deformed
quaternion as the defining matrix of a copy of $SU_q(2)\times
\b{R}^{\ge}$ ($\b{R}^{\ge}$ denoting the semigroup of nonnegative
real numbers), showing that its entries are the coordinates of
$\b{R}_q^4$. More details will be given in \cite{Fio05q}.
Then adopting the $SO_q(4)$-covariant differential
calculus on $\b{R}_q^4$ \cite{CarSchWat91} and the corresponding
Hodge duality map \cite{Fio94,Fio04JPA} in $q$-quaternion language
we have found \cite{Fio05i} solutions $A$ of the (anti)self-duality
equations, in the form of 1-form valued $2\times 2$ matrices, that
closely resemble their undeformed counterparts (instantons) in
$su(2)$ Yang-Mills theory on $\b{R}^4$. [The (still missing)
complete gauge theory might be however  a deformed $u(2)$ rather
than $su(2)$ Yang-Mills theory.] The ``coordinates of the center''
of the instanton are nevertheless noncommuting parameters,
differently from the Nekrasov-Schwarz theory. We have also found
multi-instantons solutions: they are again parametrized by
noncommuting parameters playing the role of ``size'' and
``coordinates of the center'' of the (anti)instantons. This
indicates that the moduli space of a complete theory will be a
noncommutative manifold. This is  similar to what was proposed
in \cite{IvaLecMue04} for $\b{R}_{\theta}^4$ for selfdual
deformation parameters $\theta_{\mu\nu}$.

Here we briefly report on these results.

\section{The $q$-quaternion bialgebra $C(\b{H}_q)$}

Any element $X$ in the (undeformed) quaternion algebra $\b{H}$ is given
by
\[ X=\x_1+\x_2i+\x_3j+\x_4k,
\] with $\x\in \b{R}^4$ and imaginary $i,j,k$ fulfilling
\[i^2=j^2=k^2=-1,\quad ijk=-1.
\]
Replacing $i,j,k$ by Pauli matrices $\times$ imaginary unit ${\rm
i}$,
\[ X\leftrightarrow x\equiv\left(\ba{cc}
\x_1+\x_4{\rm i} \:& \x_3+\x_2{\rm i}\\
-\x_3+\x_2{\rm i} \:& \x_1-\x_4{\rm i} \ea\right)=: \left(\ba{cc}
\alpha\:& \gamma\\
-\gamma^{\star}\:& \alpha^{\star} \ea\right) \] (where $\alpha,
\gamma\in\b{C}$), and the quaternionic product becomes represented
by matrix multiplication. Therefore  $\b{H}$ essentially consists
 of all complex $2\times 2$ matrices of this
form.

This can be $q$-deformed as follows.
We just pick the pioneering definition of the (Hopf) $*$-algebra
$C\left(SU_q(2)\right)$ \cite{Wor87a,Wor87} without imposing the det$_q$=1
condition: for $q\in\b{R}$ consider the unital associative
$\star$-algebra $\A\equiv C(\b{H}_q)$ generated by elements $\alpha,
\gamma,\alpha^{\star},\gamma^{\star}$ fulfilling the commutation
relations \be \ba{l}
\alpha\gamma=q\gamma\alpha,\qquad\alpha\gamma^{\star}=q\gamma^{\star}\alpha,
\qquad\gamma\alpha^{\star}=q\alpha^{\star}\gamma,\\[8pt]
\gamma^{\star}\alpha^{\star}=q\alpha^{\star}\gamma^{\star},\qquad
[\alpha,\alpha^{\star}]=(1\!-\!q^2)\gamma\gamma^{\star} \qquad
[\gamma^{\star},\gamma]=0.\ea \label{explqquatcomrel} \ee
Introducing the matrix
\[
x\equiv \left(\ba{ll} x^{11}\: & x^{12}\\
x^{21}\:& x^{22}\ea\right):=\left(\ba{cc} \alpha\:& -q\gamma^{\star}\\
\gamma\: & \alpha^{\star} \ea\right)
\]
 we can
rewrite these commutation relations as \be \hat R x_1x_2=x_1x_2\hat R
%\qquad\mbox{or equivalently}\qquad R^{-1}_{21}x_1x_2=x_2x_1R^{-1}_{21},
      \label{qquatcomrel}
\ee and the conjugation relations as
$x^{\alpha\beta}{}^{\star}=\epsilon^{\beta\gamma}x^{\delta\gamma}
\epsilon_{\delta\alpha}$, i.e. \be x^{\dagger}=\bar x\qquad\qquad
\mbox{where } \bar a:=\epsilon^{-1} a^T\epsilon \quad\forall a\in
M_2. \label{qquatstarrel} \ee Here we have used the braid matrix and
the $\epsilon$-tensor of $M_q(2),GL_q(2),SU_q(2)$, 
\be
\epsilon\!=\!\left(\ba{cc} 0\: & 1\\ -q \: & 0 \ea\right)\!=\!-q
\epsilon^{-1}\qquad\qquad \hat R^{\alpha\beta}_{\gamma\delta}
=q\delta^{\alpha}_{\gamma}\delta^{\beta}_{\delta}+\epsilon^{\alpha\beta}
\epsilon_{\gamma\delta} .     \label{qepsilon}  
\ee 
[with $\epsilon\!\equiv\!(\epsilon_{\alpha\beta})$ and
$\epsilon^{-1}\!\equiv\!(\epsilon^{\alpha\beta})$]. So
$\A:=C(\b{H}_q)$ can be endowed also with a bialgebra  structure (we
are not excluding the possibility that $x\equiv {\bf 0}_2$), more
precisely a real section of the bialgebra $C\left(M_q(2)\right)$ of
$2\times 2$ quantum matrices \cite{Dri86,Wor87a,FadResTak89}. Since the coproduct
$$
\Delta(x^{\alpha\gamma})=(ax)^{\alpha\gamma}
$$
is an algebra map,  the matrix product $ax$ of any two matrices
$a,x$ with mutually commuting entries and fulfilling
(\ref{qquatcomrel}-\ref{qquatstarrel}) again fulfills the latter.
Therefore we shall call any such matrix $x$ a {\it $q$-quaternion},
and $\A:=C(\b{H}_q)$ the $q$-quaternion bialgebra.

As well-known, the socalled `$q$-determinant' of $x$ \be |x|^2
\equiv\det{}_q(x):=x^{11}x^{22}-qx^{12}x^{21}=\alpha^{\star}\alpha +
\gamma^{\star}\gamma\sim x^{\alpha\alpha'}x^{\beta\beta'}
\epsilon_{\alpha\beta}\epsilon_{\alpha'\beta'}, \ee is central,
manifestly nonnegative-definite and group-like. It is zero iff
$x\equiv {\bf 0}_2$. Relations (\ref{qquatcomrel}) can be also equivalently 
reformulated as 
\be
x\bar x=\bar xx=|x|^2I_2                \label{blu}
\ee
($I_2$ denotes the unit $2\times 2$ matrix). 
If we assume $x\neq {\bf 0}_2$ and extend
$C(\b{H}_q)$ by the new (central, positive-definite) generator
$|x|^{-1}$ one finds that $x$ is invertible with inverse
\be
x^{-1}=\frac{\bar x}{|x|^2}.
\ee
$C(\b{H}_q)$ becomes a Hopf
$\star$-algebra  [a real section of $C\left(GL_q(2)\right)$]. 
The matrix elements
of $T:=\frac x{|x|}$ fulfill the relations (\ref{qquatcomrel}) and
\be T^{\dagger}=T^{-1}=\overline{T}, \qquad\qquad\det{}_q(T)=\1, \ee
namely generate as a quotient algebra $C\left(SU_q(2)\right)$
\cite{Wor87a,Wor87}, therefore in this case the entries of $x$ 
generate  the
(Hopf) $\star$-algebra of functions on the quantum group $SU_q(2)\times
GL^+(1)$, in analogy with the $q=1$ case.

As a $\star$-algebra, $\A:=C(\b{H}_q)$ coincides with the algebra of
functions on the $SO_q(4)$-covariant quantum Euclidean Space
$\b{R}_q^4$ of \cite{FadResTak89}, identifying their generators as
\be
x^1= qx^{11},\quad x^2= x^{12},\quad x^3=- qx^{21},\quad x^4= x^{22}.
                                                    \label{lintra}
\ee
The  commutation relations are preserved by the (left) coactions of
both $SO_q(4)=SU_q(2)\otimes SU_q(2)'/\b{Z}_2$
and of the extension
$\widetilde{SO_q(4)}:=SO_q(4)\!\times\! GL^+(1)=\b{H}_q\!\times\!\b{H}_q'/GL(1)$
(the quantum group of rotations and scale transformations in 4
dimensions), which take the form
\be
x\to a\, x\,b^T.
\ee
Here $a,b$ are the defining matrices of $SU_q(2),SU_q(2)'$ in the first case and of
$\b{H}_q,\b{H}_q'$
in the second (with entries commuting with each other and with those of $x$),
$b^T$ means the transpose of $b$, and matrix product is understood.

A different matrix version (with no interpretation in terms of
$q$-deformed quaternions) of a $SU_q(2)\times SU_q(2)$ covariant
quantum Euclidean space  was proposed in \cite{Maj94}.

\section{Other preliminaries}

The $SO_q(4)$-covariant {\bf differential calculus $(d,\Omega^*)$ on
$\b{R}_q^4\sim\b{H}_q$}  \cite{CarSchWat91} is obtained imposing
covariant homogeneous bilinear commutation relations
 (\ref{xxirel}) between the $x^i$ and the differentials
$\xi^i:=dx^i$.
 Partial derivatives are introduced through the
decomposition
$d=\xi^a\partial_a=\xi^{\alpha\alpha'}\partial_{\alpha\alpha'}$. All
other commutation relations are derived by consistency. The complete
list is given by \bea
&& {\sf P}_a{}^{ij}_{hk}x^hx^k=0, \label{xxrel}\\
&& x^h\xi^i=q\hat {\sf R}^{hi}_{jk}\xi^jx^k,\label{xxirel}\\
&& ({\sf P}_s+{\sf P}_t)^{ij}_{hk}\xi^h\xi^k=0,\label{xixirel}\\
&& {\sf P}_a{}^{ij}_{hk}\partial_j\partial_i=0, \label{ddrel}\\
&& \partial_i x^j = \delta^j_i+q\hat {\sf R}^{jh}_{ik}
x^k\partial_h,
                                                 \label{dxrel}\\
&& \partial^h\xi^i=q^{-1}\hat {\sf R}^{hi}_{jk}\xi^j
\partial^k.\label{dxirel} \eea $\hat{\sf R}\equiv$braid
matrix of $SO_q(4)$; ${\sf P}_s,{\sf P}_a,{\sf P}_t\equiv$
deformations of the symmetric trace-free, antisymmetric and trace
projectors appearing in the projector decomposition of $\hat{\sf
R}$. Up to the linear transformation (\ref{lintra})
$$
q\hat{\sf R}=\hat R\otimes \hat R.
$$
The Laplacian
$\Box\equiv\partial\cdot\partial:=\partial_kg^{hk}\partial_h$ is
$SO_q(4)$-invariant and commutes the $\partial_i$. In ${\cal H}$
there exists a special invertible element $\Lambda$ such that
\[
\Lambda x^i=q^{-1}x^i\Lambda,\qquad\quad
\Lambda\partial^i=q\partial^i\Lambda, \qquad\quad \Lambda
\xi^i=\xi^i\Lambda.\label{Lambdaprop}
\]

{\bf Definitions}:
\begin{itemize}

\item $\bigwedge^*\equiv$ $\natural$-graded algebra generated by the
$\xi^i$, where grading $\natural \equiv$degree in $\xi^i$; any
component $\bigwedge^p$ with $\natural =p$ carries an
irreducible representation of $U_qso(4)$ and has the same dimension
as in the $q=1$ case.

\item ${\cal DC}^*\equiv $ $\natural$-graded algebra generated by
$x^i,\xi^i,\partial_i$. Elements of ${\cal DC}^p$ are
differential-operator-valued $p$-forms.

\item $\Omega^*\equiv$ $\natural$-graded subalgebra generated by the
$\xi^i,x^i$. By definition $\Omega^0=\A$ itself, and both $\Omega^*$
and $\Omega^p$ are $\A$-bimodules. Also, we shall denote $\Omega^*$
enlarged with $\Lambda^{\pm 1}$ as $\tilde\Omega^*$, and the
subalgebra generated by $T^{\alpha\alpha'}:=x^{\alpha\alpha'}/|x|$,
$dT^{\alpha\alpha'}$ as $\Omega_S^*$
(the latter is 4-dim!).

\item ${\cal H}\equiv$subalgebra generated by the $x^i,
\partial_i$. By definition, ${\cal DC}^0={\cal H}$, and
both ${\cal DC}^*$ and ${\cal DC}^p$ are ${\cal H}$-bimodules.

\end{itemize}

The restricted (but still 4-dimensional!) differential calculus
$(\Omega_S^*,d)$ coincides with the Woronowicz 4D- on $C\left(SU_q(2)\right)$.

The special 1-form
$$ \theta:=\frac 1{1-q^{-2}}|x|^{-2}\,d|x|^2=
\frac{q^{-2}}{q^2-1}\xi^{\alpha\alpha'}\frac{x^{\beta\beta'}}{|x|^2}
\epsilon_{\alpha\beta}\epsilon_{\alpha'\beta'}
$$
plays the role of "Dirac Operator" \cite{Con94} of the differential
calculus,
\[
d\omega_p=[-\theta,\omega_p\}, \qquad\qquad \omega_p\in\Omega^p ,
\label{thetacommu}
\]

\bigskip
However, $d(f^{\star})\neq (df)^{\star}$, and moreover there is no
$\star$-structure $\star:\Omega^*\to\Omega^*$, but only a
$\star$-structure
$$
\star:{\cal DC}^*\to {\cal DC}^*
$$ \cite{OgiZum92},
with a rather nonlinear character (the latter  has been recently
\cite{Fio04} recast in a much more suggestive form).

\bigskip

The {\bf Hodge map} \cite{Fio94,Fio04JPA} is a $SO_q(4)$-covariant,
$\A$-bilinear map $*:\tilde\Omega^p\to\tilde\Omega^{4-p}$ such that
$*^2= \id$, defined by
\[ {}^*(\xi^{i_1}...\xi^{i_p})=q^{-4(p-2)} c_p\,\xi^{i_{p+1}}...\xi^{i_4}
\varepsilon_{i_4...i_{p+1}}{}^{i_1...i_p}\Lambda^{2p-4}, \] where
$\varepsilon^{hijk}\equiv$ $q$-epsilon tensor and $c_p$ are suitable
normalization factors. Actually this extends to a ${\cal H}$-bilinear map
$*:{\cal DC}^p\to{\cal DC}^{4-p}$ with the same features. For $p=2$
$\Lambda$-powers disappear and one even gets a map
$*:\Omega^2\to\Omega^2$ defined by \be {}^*\xi^i\xi^j=\frac
1{[2]_q}\xi^h\xi^k \varepsilon_{kh}{}^{ij}\omega_{ji}.
\label{defHodge2x4} \ee $\Omega^2$ (resp. ${\cal DC}^2$) splits into
the direct sum of $\A$- (resp. ${\cal H}$-) bimodules
\[
\Omega^2=\check\Omega^2\oplus \check\Omega^{2}{}' \qquad\quad
\mbox{(resp. }{\cal
DC}^2=\check{\cal DC}^2\oplus \check{\cal DC}^{2}{}'\mbox{)}
\]
of the  eigenspaces of $*$ with eigenvalues $1,-1$ respectively,
whose elements are ``self-dual and anti-self-dual 2-forms''.
$\check\Omega^2$ (resp. $\check{\cal DC}^2$) is generated by the
self-dual exterior forms $(\xi\bar\xi)^{\alpha\beta}$, 
or equivalently by the ones 
\be f^{\alpha\beta}:=
(\xi\bar\xi\epsilon)^{\alpha\beta}                      \label{deff} \ee
through (left or right) multiplication  by elements of $\A$ (resp.
${\cal H}$). $f^{\alpha\beta}$ span a (3,1) corepresentation
space of $SU_q(2)\otimes SU_q(2)'$.

One can find 1-form-valued matrices $a$ such that \be
d\,a^{\alpha\beta}=f^{\alpha\beta}; \ee $a$ is uniquely determined
to be \be a^{\alpha\beta} ={\cal P}_s{}^{\alpha\beta}_{\gamma\delta}
(\xi\epsilon x^T)^{\gamma\delta}, \label{aexplicit} \ee 
where ${\cal P}_s$ is the $SU_q(2)$-covariant symmetric projector, if we
require $a^{\alpha\beta}$ to transform as $f^{\alpha\beta}$, i.e. in
the (3,1) dimensional corepresentation of $SU_q(2)\times SU_q(2)'$,
whereas will be defined up to $d$-exact terms of the form
$$
\tilde a= a+\1_2\,dM(|x|^2)
$$  if we  just require $\tilde a^{\alpha\beta}$ to be in the
$(3,1)\oplus (1,1)$ reducible representation. In particular, the
1-form valued matrix $(dT)\overline{T}$ belongs to the latter. In the
$q=1$ limit (\ref{aexplicit}) becomes \[
a^{\alpha\beta}=\Big(\xi\epsilon x^T\Big)^{(\alpha\beta)}
=-\left\{Im(\xi\,\bar x\epsilon)\right\}^{\alpha\beta}. \]

Similarly, antiself-dual $\check \Omega^2{}'$, $\check{\cal
DC}^2{}'$ are generated by $(\bar\xi\xi)^{\alpha'\beta'}$,
or equivalently by 
\be 
f'{}^{\alpha'\beta'}:=(\bar\xi\xi\epsilon)^{\alpha'\beta'}, \label{deff'} 
\ee 
and one can find
1-forms $a'{}^{\alpha'\beta'}$ such that
$d\,a'{}^{\alpha'\beta'}=f'{}^{\alpha'\beta'}$, etc.

\bigskip
{\bf Integration over $\b{R}_q^4$} \cite{Ste96,Fio93,fiothesis}
can be introduced by the decompositon
$$
\int_{\b{R}_q^4}d^4x=\int\limits_0^{\infty}d|x| \: \int_{|x|\cdot
S_q^3}dT^3
$$
Integration over the radial coordinate
has to fulfill the scaling property
$\int\limits_0^{\infty}d|x|\,g(|x|)=\int\limits_0^{\infty}d(q|x|)\,g(q|x|)$.
Integration over the quantum sphere $S_q^3$
is determined up to normalization by the requirement of
$SO_q(4)$-invariance. The algebra of functions on  the quantum sphere
$S_q^3$ is generated by the
$T^{\alpha\beta}:=x^{\alpha\beta}/|x|$.

This integration over $\b{R}_q^4$ fulfills all the main properties
of Riemann integration over $\b{R}^4$, including Stokes' theorem,
except the cyclic property.

\section{Noncommutative gauge theories: standard framework}

The standard framework \cite{Con94,FigGraVar01,Lan97} for
noncommutative gauge theories (i.e. gauge theories on noncommutative
manifolds) closely mimics that for commutative ones. In $U(n)$ gauge
theory the gauge transformations $U$ are unitary $\A$-valued ($\A$
being the algebra of functions on the noncommutative manifold)
$n\times n$ unitary matrices, $U\!\in\! M_n(\A)\equiv
M_n(\b{C})\otimes_{\b{C}}\A$. The gauge potential $A\equiv (
A^{\dot\alpha}_{\dot\beta})$ is a antihermitean 1-form-valued
$n\times n$ matrix, $A\in M_n(\Omega^1(\A))$. The definition of the
field strength $F\in M_n(\Omega^2(\A))$ associated to $A$ is as
usual $F:=dA+AA$. At the right-hand side the product $AA$ has to be
understood both as a (row by column) matrix product and as a wedge
product. Even for $n=1$, $AA\neq 0$, contrary to the commutative
case. The Bianchi identity $DF:=dF+[A,F]=0$ is automatically
satisfied and the Yang-Mills equation reads as usual $D{}^*F=0$.
Because of the Bianchi identity, the latter is automatically
satisfied by any solution of the (anti)self-duality equations \be
{}^*F=\pm F. \ee

The Bianchi identity, the  Yang-Mills equation, the
(anti)self-duality equations,  the
flatness condition $F=0$ are preserved by gauge transformations
$$
A^U=U^{-1}(AU +dU), \qquad\Rightarrow \qquad F^U= U^{-1}F U.
$$
As usual, $A=U^{-1}dU$ implies $F=0$. Up to normalization factors,
the gauge invariant `action' $S$ and `Pontryagin index' $\c{Q}$
are defined by \be
S = \mbox{Tr}(F\:{}^*\!F), \qquad\qquad\qquad  \c{Q} = \mbox{Tr}(FF)
              \label{actionfun} \ee where Tr stands for  a
positive-definite trace combining the $n\times n$-matrix trace with
the integral over the noncommutative manifold (as such, Tr has to
fulfill the cyclic property). If integration $\int$ fulfills itself
the cyclic property then this is obtained by simply choosing $
\mbox{Tr}=\int  \mbox{tr}$, where $\mbox{tr}$ stands for the
ordinary matrix trace. $S$ is automatically nonnegative.

\bigskip

In the present $\A\equiv C(\b{R}_q^4)=C(\b{H}_q)$ case ther are {\bf 2 main
problems}:

\begin{enumerate}

\item Integration over $\b{R}_q^4$ fulfills a {\it deformed} cyclic property
\cite{Ste96}.

\item $d(f^{\star})\neq (df)^{\star}$, and there is no
$\star$-structure $\star:\Omega^*\to\Omega^*$, but only a
$\star$-structure  $\star:{\cal DC}^*\to {\cal DC}^*$ \cite{OgiZum92},
with a nonlinear character.

\end{enumerate}

A solution to both problems might be obtained
\begin{enumerate}

\item allowing for ${\cal DC}^1$-valued $A$ ($\Rightarrow$
${\cal DC}^2$-valued $F$'s), and/or

\item realizing Tr$(\cdot)$
by in the form $\mbox{Tr}(\cdot)$:=$\int\mbox{tr}(W\cdot)$, with $W$ some
suitable positive definite ${\cal H}$-valued (i.e.
pseudo-differential-operator-valued) $n\times n$ matrix (this
implies a change in the hermitean conjugation of differential
operators), or even a more general form.

\end{enumerate}

This hope is based on our results \cite{Fio04}: 1) the
$\star$-structure  $\star:{\cal DC}^*\to {\cal DC}^*$ can been
recast in a more suggestive form of similarity transformations
(involving the realization as pseudodifferential operators of the
ribbon element $\tilde w$ and of the "vector field generators"
$\tilde Z^i_j$ of the central extension of $U_qso(4)$ with
dilatations); 2)
 $d$ and the exterior coderivative $\delta:=-*d*$
become conjugated of each other $$ (\alpha_p, d\beta_{p\!-\!1}) =
(\delta\alpha_p, \beta_{p\!-\!1}), \qquad\qquad (
d\beta_{p\!-\!1},\alpha_p) = (\beta_{p\!-\!1},\delta\alpha_p) $$ if
one defines
$$
(\alpha_p, \beta_p)=\int_{\b{R}_q^4}
\alpha_p^{\star}\:\:{}^*\:\tilde w'{}^{1/2}\beta_p
$$
where $\tilde w'$ is the realization of $\tilde w$ as a
pseudodifferential operator.

\section{The (anti)instanton solution}

 We first recall the {\it commutative ($q\!=\!1$)
solution} of the self-duality eq. ${}^*F=F$: the
instanton solution of \cite{BelPolSchTyu75} in t' Hooft
\cite{tHo76} and in ADHM \cite{AtiDriHitMan78}  quaternion notation
(see \cite{Ati79} for an introduction) reads: \bea A &=&  dx^i\, \sigma^a\,
\underbrace{\eta^a_{ij}x^j \frac 1{\rho^2+r^2/2}}_{A^a_i},\nn
&=&-Im\left\{\xi\,\frac{\bar x}{|x|^2}\right\} \frac
1{1+{\rho^2}\frac 1{|x|^2}}\nn  &=& -(dT)\overline{T} \frac
1{1+{\rho^2}\frac 1{|x|^2}}\label{inst}\\
 F &=& \xi\bar\xi\,\rho^2\frac
1{(\rho^2+|x|^2)^2}, \label{Finst} \eea where $r^2:=x\cdot
x=2|x|^2$, $\eta^a_{ij}$ are the so-called 't Hooft $\eta$-symbols and $\rho$
is the size of the instanton (here centered at the origin).
The third equality is based on the identity
$$
\xi\,\frac{\bar x}{|x|^2}=(dT)\overline{T}+I_2\frac {d|x|^2}{2|x|^2}
$$
and the observation that the first and second term at the rhs are
respectively antihermitean and hermitean,  i.e. the imaginary and
the real part of the quaternion at the lhs.

\bigskip
{\it Noncommutative ($q\!\neq \!1$) solutions} of ${}^*F=F$. Looking for
$A$ directly in the form  $A=\xi\bar x\,l/|x|^2+\theta\, I_2\,n$,
where $l,n$ are functions of $x$ only through $|x|$,
one finds a family of solutions
parametrized by  $\rho^2$ (a nonnegative constant, or more generally a further
generator of the algebra) and by the function $l$ itself.
The freedom in the choice of $l$ should disappear upon
imposing the proper (and still missing) antihermiticity
condition on $A$, as it occurs in the $q=1$ case. For the moment, out of this
large family we just pick one which has the right $q\to 1$ limit
and closely resembles the undeformed solutions (\ref{inst}-\ref{Finst}).
\be
\ba{l} A = - (dT)\overline{T} \,\frac 1{1+{\rho^2}\frac
1{|x|^2}},\\[8pt] F = q^{-1}\xi\bar\xi \frac 1{|x|^2+\rho^2}
\rho^2\frac 1{q^2|x|^2+\rho^2}.\ea\label{qinst}
\ee
Of course we have to extend the algebras so that they contain the rational
functions at the rhs.
The matrix elements $A^{\alpha\beta}$  span a
$(3,1)\oplus (1,1)$ dimensional corepresentation of
$SU_q(2)\times SU_q(2)'$,
suggesting as the `fiber' of
the gauge group  in the complete theory
a (possibly deformed) $U(2)$ [instead of a $SU(2)$].

One can shift the `center of the instanton' away from the origin by
the replacement (or `braided coaddition' \cite{Maj95})
$$
x\to x-y,
$$
where the `coordinates of the center' $y^i$ generate a new copy of
$\A$, `braided' with the original one (see below).
Therefore the instanton moduli space must be a noncommutative manifold, with
coordinates $\rho, y^i$! This is similar to what was proposed
in \cite{IvaLecMue04} for the instanton moduli space  on $\b{R}_{\theta}^4$.

By the scaling and translation invariance of integration over
$\b{R}_q^4$,  if we could find a `good' pseudodifferential operator
$W$  to define gauge invariant ``action'' and ``topological charge''
by $$ \c{Q}:=\int_{\b{R}_q^4}\mbox{tr}(WF\,F)=
\int_{\b{R}_q^4}\mbox{tr}(W F\,{}^*F)=S $$ the latter would, as in
the commutative case, equal a constant independent of $\rho,y$
(which by the choice of the normalization of the integral we can
make 1).

In the $q=1$ case multi-instanton solution are explicitly written
down in the socalled `singular gauge'. Note that as in the $q=1$
case $T=x/|x|$ is unitary and singular at $x=0$. So it can play the
role of a `singular gauge transformation'. In fact $A$ can be
obtained through the gauge transformation $A=T(\hat
A\overline{T}+d\overline{T})$ from the singular gauge potential \bea
\hat A &=& \overline{T}dT\frac 1{1+|x|^2\frac 1{\rho^2}} =-\frac
1{1+|x|^2\frac 1{q^2\rho^2}}(d\overline{T})T\nn &=&-\frac
1{1+|x|^2\frac 1{q^2\rho^2}} \left[q^{-1}\bar\xi\frac x{|x|^2}-
\frac {q^{-3}I_2}{1\!+\!q}\left(
\xi^{\alpha\alpha'}\frac{x^{\beta\beta'}}{|x|^2}
\epsilon_{\alpha\beta}\epsilon_{\alpha'\beta'}\right)\right]. \label{hatA}
\eea 
$\hat A$ can be expressed also in
the form \[
\hat A= \phi^{-1}\hat{\cal D}\phi, \qquad\qquad \phi:=1+q^2\rho^2\frac
1{|x|^2},
\]
where $\hat{\cal D}$ is the first-order-differential-operator-valued
$2\times 2$ matrix obtained from the square bracket in (\ref{hatA}) by the
replacement $x^{\alpha\alpha'}/|x|^2\to q^2\partial^{\alpha\alpha'}$:
\be
\hat{\cal D}:=q\bar\xi\partial-\frac {q^{-1}I_2}{q\!+\!1}d \label{Doper}
\ee
(for simplicity we are here assuming that $\rho^2$ commutes with
$\xi^{\alpha\alpha'}\partial^{\beta\beta'}$). $\phi$ is harmonic:
$$
\Box\phi=0.
$$
This is the analog of the $q=1$ case, and is useful for the
construction of multi-instanton solutions.

\bigskip

The {\bf anti-instanton solution} is obtained just by converting
unbarred into barred matrices, and conversely, as in the $q=1$ case.
For instance, from (\ref{qinst}) we obtain the anti-instanton
solution in the regular gauge \be \ba{l} A' = - (d\overline{T})T
\,\frac 1{1+{\rho^2}\frac
1{|x|^2}},\\[8pt] F' = q^{-1}\bar\xi\xi \frac 1{|x|^2+\rho^2}
\rho^2\frac 1{q^2|x|^2+\rho^2}.\ea\label{qantiinst}
\ee

\section{Multi-instanton solutions}

We have found solutions of the self-duality equation corresponding to
$n$ instantons in the ``singular
gauge''  \cite{tHo76,vari} in the form
\be
\hat A= \phi^{-1}
\hat{\cal D}\phi,
\ee
where $\phi$ is the harmonic scalar function
\be
\phi=1+\rho_1^2\frac 1{(x\!-\!y_1)^2}+
\rho_2^2\frac 1{(x\!-\!y_1\!-\!y_2)^2}+...+\rho_n^2\frac 1{(x\!-\!y_1\!-\!...\!-\!y_n)^2}
\ee
as in the commutative case. In the commutative limit
\bea
&& \rho_{\mu}\equiv \mbox{size of the $\mu$-th instanton},\nn
&& v^i_{\mu}:=\sum\limits_{\nu=1}^{\mu}y^i_{\nu}
\equiv \mbox{$i$-th coordinate of the $\mu$-th instanton}.\nonumber
\eea
are constants ($\mu=1,2,...,n$).
In the noncommutative setting the new generators
$\rho_{\mu}^2,y^i_{\nu}$ have to fulfill the following nontrivial
commutation relations: \bea
&&\rho_{\nu}^2\rho_{\mu}^2=q^2\,\rho_{\mu}^2\rho_{\nu}^2\qquad
\qquad \nu<\mu             \nn
&&\rho_{\nu}^2y_{\mu}^i=y_{\mu}^i\rho_{\nu}^2 \cdot \left\{\ba{l}
q^{-2}\:\: \nu<\mu \cr 1, \:\:\nu \ge\mu \ea\right. \nn
&&\rho_{\mu}^2\xi^i=\xi^i\rho_{\mu}^2, \qquad\qquad
\partial_i\rho_{\mu}^2=\rho_{\mu}^2\partial_i.\label{ntcr}\\
&& y_{\mu}^iy_{\nu}^j=q\RH^{ij}_{hk}y_{\nu}^hy_{\mu}^k\qquad\qquad
\nu<\mu, \nn &&\PH_A{}^{ij}_{hk}y_{\mu}^hy_{\mu}^k=0. \nonumber
\eea
($\mu,\nu=0,1,...,n$, and we have set $x^i\equiv y_0^i$).

The last relation states that for any fixed $\nu$ the 4 coordinates
$y_{\nu}^i$ generate a copy of $\A$. The last but one states that
the various copies of $\A$ are {\it braided} \cite{Maj95} w.r.t.
each other (this is necessary for the $SO_q(4)$ covariance of the
overall algebra).

\medskip
The obvious consequence of the nontrivial
commutation relations (\ref{ntcr}) is that in a complete theory
{\bf the instanton moduli space must be a noncommutative manifold}.

\medskip
At least for low $n$, we have been able to go to a gauge potential $A$
`regular'  in $z_{\mu}^i\!:=\!x^i\!-\!v^i_{\mu}$
by a `singular gauge transformation' (as in the $q=1$ case
\cite{GiaRot77,OliSciCre79,vari}), which also depends on
$\rho_{\mu}^2,y^i_{\nu}$.

\end{document}